\documentclass[aps,prd,onecolumn,groupedaddress,showpacs,nofootinbib,amssymb]{revtex4}
\usepackage[dvips]{graphicx,color}
\usepackage{amssymb}
\usepackage{amsmath}
\usepackage{graphicx}
\usepackage{amsfonts}
\usepackage{bm}
\usepackage{cancel}

\newcommand{\be}{\begin{equation}}
\newcommand{\ee}{\end{equation}}
\newcommand{\bea}{\begin{eqnarray}}
\newcommand{\eea}{\end{eqnarray}}
\newcommand{\nn}{\nonumber \\}
\newcommand{\e}{\mathrm{e}}

\newcommand{\br}{\bm{r}}

\begin{document}

\title{The bounce universe history from unimodular $F(R)$ gravity}
\author{S.~Nojiri,$^{1,2}$\,\thanks{nojiri@gravity.phys.nagoya-u.ac.jp}
S.~D.~Odintsov,$^{3,4}$\,\thanks{odintsov@ieec.uab.es}
V.~K.~Oikonomou,$^{5,6}$\,\thanks{v.k.oikonomou1979@gmail.com}}
\affiliation{$^{1)}$ Department of Physics, Nagoya University, Nagoya 464-8602,
Japan \\
$^{2)}$ Kobayashi-Maskawa Institute for the Origin of Particles and the
Universe, Nagoya University, Nagoya 464-8602, Japan \\
$^{3)}$Institut de Ciencies de lEspai (IEEC-CSIC),
Campus UAB, Carrer de Can Magrans, s/n\\
08193 Cerdanyola del Valles, Barcelona, Spain\\
$^{4)}$ ICREA, Passeig LluAs Companys, 23,
08010 Barcelona, Spain\\
$^{5)}$ Tomsk State Pedagogical University, 634061 Tomsk, Russia\\
$^{6)}$ Laboratory for Theoretical Cosmology, Tomsk State University of Control Systems
and Radioelectronics (TUSUR), 634050 Tomsk, Russia\\
}

\begin{abstract}
In this paper we investigate how to realize various quite well known cosmological bouncing models in the context of the recently developed unimodular $F(R)$ gravity. Particularly, we shall study the matter bounce scenario, the singular bounce, the superbounce and a symmetric bounce scenario. We present the behavior of the Hubble radius for each of the bouncing models we shall take into account and we investigate which era of the bouncing model is responsible for the cosmological perturbations. As we shall demonstrate, the various bouncing models do not behave in the same way, so the cosmological perturbations for each model may correspond to a different era, in comparison to other models. Also we present which unimodular $F(R)$ gravity realizes each model. We also show that Newton's law is not modified in the unimodular $F(R)$ gravity, which also is proven to be a ghost-free theory, and in addition we discuss the matter stability issue. Finally, we demonstrate how it is possible to solve a cosmological constant problem in the context of unimodular $F(R)$ gravity. 
\end{abstract}


\maketitle

\section{Introduction}

The Big Bounce alternative to the standard Big Bang theory is a quite appealing scenario, since the initial singularity problem is consistently remedied. In the context of the bouncing cosmology, the Universe contracts until a minimal radius is reached, at which point the Universe bounces off and starts to expand. Since a minimal radius is reached, there is no initial singularity, as in the standard inflationary cosmologies, and this is a quite appealing feature. Apart from the absence of the initial singularity, it has been shown in the literature that the Big Bounce cosmologies \cite{Novello:2008ra,Li:2014era,Cai:2013kja,Cai:2014zga,deHaro:2014kxa,Qiu:2010vk,Qiu:2010ch,Lehners:2011kr,Easson:2011zy,Cai:2012va,Odintsov:2015zua,Cai:2007qw,
Khoury:2001bz,Erickson:2003zm,Cai:2011tc,Koehn:2013upa,Odintsov:2015uca,Oikonomou:2014yua,
Brandenberger:2012zb,Quintin:2014oea,Cai:2011ci,Cai:2011zx,Bamba:2012ka,deHaro:2012xj,deHaro:2014jva,deHaro:2015wda,Bamba:2013fha,Barragan:2009sq} can provide a viable and consistent alternative scenario to the standard inflationary paradigm (for reviews see for instance \cite{Mukhanov:2005sc,Gorbunov:2011zzc,Linde:2014nna,Lyth:1998xn,Bamba:2015uma}). In principle, someone could claim that the inflationary paradigm is more likely to have occurred in the past, but this is not so, since the latest observations from the Planck collaboration \cite{Ade:2015lrj,Planck:2013jfk} strongly indicate that the resulting power spectrum of the primordial curvature perturbations is almost scale invariant, which cannot be necessarily connected to an inflationary era. Of course, many inflationary models can result to a nearly scale invariant power spectrum, but this is not a proof that inflation ever existed. In fact, it is possible that bouncing cosmologies can generate a nearly scale invariant power spectrum too, as in the case of the matter bounce scenario \cite{Brandenberger:2012zb,Quintin:2014oea,Cai:2011ci,Cai:2011zx,Bamba:2012ka,deHaro:2012xj,deHaro:2014jva,deHaro:2015wda,Bamba:2013fha,Barragan:2009sq}. In view of the cosmological viability of the bouncing cosmologies, and also due to the appealing feature of not having the initial singularity problem, in this paper we shall investigate how bouncing cosmologies can be realized by the recently developed unimodular $F(R)$ gravity theory \cite{Nojiri:2015sfd}.

Initially, the unimodular Einstein-Hilbert gravity \cite{Anderson:1971pn,Buchmuller:1988wx,Henneaux:1989zc,Unruh:1988in,Ng:1990xz,Finkelstein:2000pg,Alvarez:2005iy,
Alvarez:2006uu,Abbassi:2007bq,Ellis:2010uc,Jain:2012cw,Singh:2012sx,Kluson:2014esa,Padilla:2014yea,Barcelo:2014mua,
Barcelo:2014qva,Burger:2015kie,Alvarez:2015sba,Jain:2012gc,Jain:2011jc,Cho:2014taa,Basak:2015swx,Gao:2014nia,
Eichhorn:2015bna,Saltas:2014cta,Nojiri:2016mlb} theory was developed in order to provide a solution to the cosmological constant problem \cite{Peebles:2002gy,Nojiri:2016mlb}, in which case the cosmological constant arises from the trace-free part of the Einstein equations, and effectively the cosmological constant is not added by hand in the theory. In the standard unimodular Einstein-Hilbert gravity, the technique for obtaining a trace-free part of the Einstein equations is to fix the determinant of the metric $\sqrt{-g}$ to be a fixed number, or some function of the space-time coordinates used. An interesting feature of the resulting theory is that the cosmological perturbations are the same with the Einstein-Hilbert gravity, at least at the linear perturbation theory approach \cite{Basak:2015swx,Gao:2014nia}. Nevertheless, it is possible to spot some differences on the microwave temperature anisotropies and gravitational potential Sachs-Wolfe relation, since in the unimodular gravity, the shift variable cannot be equal to zero, as was demonstrated in Ref.~\cite{Gao:2014nia}.

In a recent study \cite{Nojiri:2015sfd} we extended the unimodular gravity formalism in the case of $F(R)$ gravity (for reviews on $F(R)$ gravity, see \cite{Nojiri:2010wj,Nojiri:2013zza,Nojiri:2006ri,Capozziello:2010zz,Capozziello:2011et,Bamba:2012cp}), in which case, in order for the unimodular constraint to be satisfied, we modified the metric accordingly. Also we introduced a Lagrange multiplier, which by varying the metric with respect to it, resulted to the unimodular constraint. Our motivation was the fact that in the context of the standard $F(R)$ gravity, many exotic scenarios which was not possible to realize in the standard Einstein-Hilbert gravity, can be consistently realized in the $F(R)$ gravity theory, for example the unified description of early and late-time acceleration \cite{Nojiri:2003ft}. For more viable examples of $F(R)$ gravity unifying inflation with dark energy, see Refs. \cite{Dunsby:2010wg,Hu:2007nk,Capozziello:2005ku,Faraoni:2006sy,Olmo:2005zr,Appleby:2007vb,
Barrow:2005dn,add1,add2}. As a result of our work, we provided a reconstruction method which enabled us to realize various cosmological scenarios. In this paper we shall use the formalism we developed in \cite{Nojiri:2015sfd}, in order to investigate how to realize certain bouncing cosmologies. Particularly, we shall investigate how the matter bounce scenario \cite{Brandenberger:2012zb,Quintin:2014oea,Cai:2011ci,Cai:2011zx,Bamba:2012ka,deHaro:2012xj,deHaro:2014jva,
deHaro:2015wda,Bamba:2013fha,Barragan:2009sq}, the superbounce \cite{Koehn:2013upa,Odintsov:2015uca,Oikonomou:2014yua}, the singular bounce \cite{Odintsov:2015ynk,Oikonomou:2015qfh,Oikonomou:2015qha,Odintsov:2015zza} and the symmetric bounce can be realized in the context of a vacuum unimodular $F(R)$ gravity theory, and with vacuum we mean that no matter fluids are present. After providing some essential features for the aforementioned cosmologies, we proceed by using the reconstruction method \cite{Nojiri:2015sfd}, and we present which unimodular $F(R)$ gravity can realize the aforementioned cosmologies. 

Another issue that we will address, is related to the cosmological constant in the context of the unimodular $F(R)$ gravity. Using some simple considerations, we demonstrate how a cosmological constant can naturally arise in the context of the unimodular $F(R)$ gravity, as in the case of the unimodular Einstein-Hilbert gravity. 

Finally, in all the studies which we present in the next sections, we assumed that the bounce occurs in the cosmological time $t$-variable. The unimodular $F(R)$ gravity is related to another coordinate $\tau$, related to the cosmological time $t$. We shall call the metric containing the variable $\tau$, as the $\tau$-variable. Then, in addition to the $t$-variable studies we shall perform, in principle the same calculations can be done in the $\tau$-variable, that is, by assuming that the cosmological bounce occurs in the $\tau-$variable. It can be easily shown that in general, the results in the two aforementioned cases are different, except for the cases in which the $\tau$ and $t$ variables are trivially related, that is, the variable $\tau$ depends linearly on the variable $t$. 

As a general remark, the resulting picture of the unimodular $F(R)$ gravity, is different from the ordinary $F(R)$ gravity description, owing to the existence of the Lagrange multiplier, and as we demonstrate, this can be seen at a quantitative level, by computing corrections to Newton's law, and when we discuss the stability conditions.

This paper is organized as follows: In section II we present the fundamental features of the unimodular $F(R)$ formalism and of the corresponding reconstruction mechanism, in order to render the article self-contained. In section III, by using some simple theoretical arguments, we investigate how a cosmological constant can arise in the context of the unimodular $F(R)$ gravity. In section IV we study some implications of the unimodular $F(R)$ gravity to Newton's law and to the matter stability issue and also we discuss which are the propagating modes and the possibility that a ghost mode exists. As we demonstrate, the Newtonian potential is not affected in the context of the unimodular $F(R)$ gravity, in contrast to the ordinary $F(R)$ gravity approach, where new terms appear. Also the only propagating mode is the graviton and no ghost appears. However, the matter stability issue cannot be addressed by using the standard approach used in ordinary $F(R)$ gravity. In section V, we briefly discuss the qualitative features of some well known bouncing models, and also we present which unimodular $F(R)$ gravity can realize these cosmologies. Finally, the conclusions follow in the end of the paper.

\section{The Unimodular $F(R)$ Gravity Formalism}

In order to render the article self-contained, in this section we briefly review the fundamental features of unimodular $F(R)$ gravity, but a more detailed presentation can be found in \cite{Nojiri:2015sfd}. The unimodular $F(R)$ gravity approach is based on the assumption that the metric satisfies the following constraint,
\be
\label{Uni1}
\sqrt{-g}=1\, ,
\ee
to which we shall refer to as the unimodular constraint hereafter, for simplicity. In addition, we assume that the metric expressed in terms of the cosmological time $t$ is a flat Friedman-Robertson-Walker (FRW) of the form,
\be
\label{FRW}
ds^2 = - dt^2 + a(t)^2 \sum_{i=1}^3 \left( dx^i \right)^2 \, .
\ee
The metric (\ref{FRW}) does not satisfy the unimodular constraint (\ref{FRW}), and in \cite{Nojiri:2015sfd} in order to tackle with this problem, we redefined the cosmological time $t$, to a new variable $\tau$, as follows,
\be
\label{Uni2}
d\tau = a(t)^3 dt\, ,
\ee
in which case, the metric of Eq.~(\ref{FRW}), becomes the ``unimodular metric'',
\be
\label{UniFRW}
ds^2 = - a\left(t\left(\tau\right)\right)^{-6} d\tau^2 + a\left(t\left(\tau\right)\right)^{2}
\sum_{i=1}^3 \left( dx^i \right)^2 \, ,
\ee
and hence the unimodular constraint is satisfied. Assuming the unimodular metric of Eq.~(\ref{UniFRW}), by making use of the Lagrange multiplier method \cite{Lim:2010yk,Capozziello:2013xn,Capozziello:2010uv}, the vacuum Jordan frame unimodular $F(R)$ gravity action is,
\be
\label{Uni11}
S = \int d^4 x \left\{ \sqrt{-g} \left( F(R) - \lambda \right) + \lambda \right\} \, ,
\ee
with $F(R)$ being a suitably differentiable function of the Ricci scalar $R$, and $\lambda$ stands for the Lagrange multiplier function. Note that we assumed that no matter fluids are present and also if we vary the action (\ref{Uni11}) with respect to the function $\lambda $, we obtain the unimodular constraint (\ref{Uni1}). In the metric formalism, the action is varied with respect to the metric, so by doing the variation, we obtain the following equations of motion,
\be
\label{Uni12}
0=\frac{1}{2}g_{\mu\nu} \left( F(R) - \lambda \right) - R_{\mu\nu} F'(R)
+ \nabla_\mu \nabla_\nu F'(R) - g_{\mu\nu}\nabla^2 F'(R) \, .
\ee
By using the metric of Eq.~(\ref{UniFRW}), the non-vanishing components of the Levi-Civita connection in terms of the scale factor $a(\tau)$ and of the generalized Hubble rate $K(\tau)=\frac{1}{a} \frac{da}{d\tau}$, are given below,
\be
\label{Uni13}
\Gamma^{\tau}_{\tau \tau} = - 3 K\, , \quad \Gamma^t_{ij} = a^8 K \delta_{ij}\, , \quad
\Gamma^i_{jt} = \,\, \Gamma^i_{\tau j} = K \delta_j^{\ i} \, .
\ee
The non-zero components of the Ricci tensor are,
\begin{equation}
\label{nontrivilar}
R_{\tau \tau} = - 3 \dot K - 12 K^2\, , \quad
R_{ij} = a^8 \left( \dot K + 6 K^2 \right) \delta_{ij}\, .
\end{equation}
while the Ricci scalar $R$ is the following,
\begin{equation}
\label{scalarunifrw}
R = a^6 \left( 6 \dot K + 30 K^2 \right) \, .
\end{equation}
The corresponding equations of motion become,
\begin{align}
\label{Uni14}
0 = & - \frac{a^{-6}}{2} \left( F(R) - \lambda \right) + \left( 3 \dot K + 12 K^2 \right) F'(R)
 - 3 K \frac{d F'(R)}{d\tau} \, , \\
\label{Uni15}
0 = & \frac{a^{-6}}{2} \left( F(R) - \lambda \right) - \left( \dot K + 6 K^2 \right) F'(R)
+ 5 K \frac{d F'(R)}{d\tau} + \frac{d^2 F' (R)}{d\tau^2}  \, ,
\end{align}
with the ``prime'' and ``dot'' denoting as usual differentiation with respect to  the Ricci scalar and $\tau$, respectively. Equations ~(\ref{Uni14}) and (\ref{Uni15}) can be further combined to yield the following equation,
\be
\label{Uni16}
0 = \left( 2 \dot K + 6 K^2 \right) F'(R)
+ 2 K \frac{d F'(R)}{d\tau} + \frac{d^2 F' (R)}{d\tau^2} + \frac{a^{-6}}{2} \, .
\ee
Basically, the reconstruction method for the vacuum unimodular $F(R)$ gravity, which we proposed in \cite{Nojiri:2015sfd} is based on Eq.~(\ref{Uni16}), which when it is solved it yields the function $F' = F'(\tau)$. Correspondingly, by using Eq.~(\ref{scalarunifrw}), we can obtain the function $R=R(\tau)$, when this is possible so by substituting back to $F' = F'(\tau)$ we obtain the function $F'(R)=F'\left( \tau\left(R\right) \right)$. Finally, the function $\lambda (\tau)$ can be found by using Eq.~(\ref{Uni14}), and substituting the solution of the differential equation (\ref{Uni16}). Based on the reconstruction method we just presented, we demonstrate how some important bouncing cosmologies can be realized. Note that the bouncing cosmologies shall be assumed to be functions of the cosmological time $t$, so effectively this means that the bounce occurs in the $t$-dependent FRW metric of Eq.~(\ref{FRW}). In a later section we study the scenario in which the bounce occurs in the $\tau$-variable of the metric (\ref{UniFRW}).

\section{Cosmological Constant Solution from Unimodular $F(R)$ Gravity}

Current observational data indicate that the Universe is expanding in an accelerating way, and the expansion is generated by the energy density, whose magnitude is about $\left( 10^{-3}\, \mathrm{eV} \right)^4$. On the other hand, we know that in the quantum field theory description, the quantum correction to the energy density, which is called the vacuum energy $\rho_\mathrm{vacuum}$, corresponding to matter fields, diverges and we need to introduce a cutoff scale $\Lambda_\mathrm{cutoff}$, so that the energy density is equal to, 
\be
\label{01}
\rho_\mathrm{vacuum} = \frac{1}{\left( 2 \pi \right)^3}\int d^3 k \frac{1}{2} \sqrt{ k^2 + m^2 } \sim \Lambda_\mathrm{cutoff}^4 \, .
\ee
The energy density appearing in Eq.~(\ref{01}), is much larger than the observed value $\left( 10^{-3}\, \mathrm{eV} \right)^4$, even if supersymmetry is restored in the high energy regime, in which case the quantum corrected vacuum energy is equal to, 
\begin{equation}\label{au}
\sim \Lambda_{\mathrm{cutoff}^2} \Lambda_{\cancel{\mathrm{SUSY}}}^2\, ,
\end{equation}
where with $\Lambda_{\cancel{\mathrm{SUSY}}}$ we denoted the mass scale of the supersymmetry breaking. In fact, it holds true that the energy density is equal to,
\be
\label{02}
\rho_\mathrm{vacuum} = \frac{1}{\left( 2 \pi \right)^3}\int d^3 k \frac{1}{2} \left( \sqrt{ k^2 + m_\mathrm{boson}^2 } 
 - \sqrt{ k^2 + m_\mathrm{fermion}^2 } \right)  
\sim \Lambda_\mathrm{cutoff}^2 \Lambda_{\cancel{\mathrm{SUSY}}}^2 \, ,
\ee
where we have assumed that the scale of supersymmetry breaking is given by the difference between the masses of bosons and fermions, $\Lambda_{\cancel{\mathrm{SUSY}}}^2 = m_\mathrm{boson}^2 - m_\mathrm{fermion}^2$. If we make use of a counter-term in order to obtain a very small vacuum energy of the order, $\left(10^{-3}\, \mathrm{eV}\right)^4$, we need to fine-tune this counter-term to a great extent, and this is extremely unnatural. These considerations strongly indicate our lack of understanding the quantum theory of gravity, the lack of a consistent quantum theory of gravity. Recently an interesting mechanism to make the magnitude of the vacuum energy much smaller and consistent with the observations, was proposed, see Refs.~\cite{Kaloper:2013zca,Kaloper:2014dqa,Kaloper:2015jra}, which may be called as sequestering models. In the first paper \cite{Kaloper:2013zca}, the proposed action has the following form: 
\be
\label{0Vcm1}
S = \int d^4 x \sqrt{-g } \left\{ \frac{R}{2\kappa^2} - \Lambda + \e^{2\sigma} \mathcal{L}_\mathrm{matter} \left( \e^\sigma g_{\mu\nu}, \varphi \right) \right\}
 - F\left( \frac{ \Lambda \e^{-2\sigma}}{\mu^4} \right)\, .
\ee
In Eq.~(\ref{0Vcm1}), the term $F$ is an adequate function and $\mu$ is a parameter with the dimension of mass, while $\mathcal{L}_\mathrm{matter}$ is the Lagrangian density of the matter fields present. We should note that $\Lambda$ and $\varphi$ are dynamical variables which do not depend on the spacetime coordinates. In Ref.~\cite{Kaloper:2013zca}, the functions $\e^{\frac{\sigma}{2}}$ and $F$ are denoted by $\lambda$ and $\sigma$. The variation of the action of 
Eq.~(\ref{0Vcm1}) with respect to $\Lambda$ yields, 
\be
\label{0Vcm3}
\mu^4 F' \left( \mu^4 \Lambda \e^{-2\sigma} \right) = - \e^{2\sigma}\int d^4 x \sqrt{-g}  \, .
\ee
On the other hand, by varying the action of Eq.~(\ref{0Vcm1}) with respect to $\sigma$, we obtain,
\be
\label{0Vcm4}
\int d^4 x \sqrt{-g} \e^{-\sigma} g^{\mu\nu} T\left( \e^\sigma g_{\mu\nu}, \varphi \right)_{\mu\nu} 
=   - 4 \mu^4 \Lambda \e^{-2\sigma} F' \left( \Lambda \e^{-2\sigma} \right) \, .
\ee
Note that $T\left( \e^\sigma g_{\mu\nu}, \varphi \right)_{\mu\nu}$ is the energy-momentum tensor coming from the matter fluids present, including the quantum corrections. By combining Eqs.~(\ref{0Vcm3}) and (\ref{0Vcm4}), we find
\be
\label{0Vcm5}
\left< \e^{-\sigma} g^{\mu\nu} T\left( \e^\sigma g_{\mu\nu}, \varphi \right)_{\mu\nu} \right> = 4 \Lambda \, .
\ee
Note that $\left< \e^{-\sigma} g^{\mu\nu} T\left( \e^\sigma g_{\mu\nu}, \varphi \right)_{\mu\nu} \right>$ expresses the average of $\e^{-\sigma} g^{\mu\nu} T\left( \e^\sigma g_{\mu\nu}, \varphi \right)_{\mu\nu}$ with respect to the spacetime metric,
\be
\label{03}
\left< \e^{-\sigma} g^{\rho\sigma} T\left( \e^\sigma g_{\mu\nu}, \varphi \right)_{\rho\sigma} \right> \equiv \frac{\int d^4 x \sqrt{-g} \e^{-\sigma} g^{\mu\nu} T\left( \e^\sigma g_{\mu\nu}, \varphi \right)_{\mu\nu}}{\int d^4 x \sqrt{-g}} \, .
\ee
By varying with respect to the metric we get,
\be
\label{0Vcm6}
0= - \frac{1}{2\kappa^2}\left( R_{\mu\nu} - \frac{1}{2} R g_{\mu\nu}\right) 
 - \frac{1}{2} \Lambda g_{\mu\nu}
+ \frac{1}{2}T\left( \e^\sigma g_{\mu\nu}, \varphi \right)_{\mu\nu} \, .
\ee
By using the condition of Eq.~(\ref{0Vcm5}), we may rewrite Eq.~(\ref{0Vcm6}) as follows,
\be
\label{0Vcm7}
0= - \frac{1}{2\kappa^2}\left( R_{\mu\nu} - \frac{1}{2} R g_{\mu\nu}\right) 
 - \frac{1}{8} \left< \e^{-\sigma} g^{\rho\sigma} T\left( \e^\sigma g_{\mu\nu}, \varphi \right)_{\rho\sigma} \right> g_{\mu\nu}
+ \frac{1}{2}T\left( \e^\sigma g_{\mu\nu}, \varphi \right)_{\mu\nu}\, .
\ee
In the combination $ - \frac{1}{8} \left< \e^{-\sigma} g^{\rho\sigma} T\left( \e^\sigma g_{\mu\nu}, \varphi \right)_{\rho\sigma} \right> g_{\mu\nu}
+ \frac{1}{2}T\left( \e^\sigma g_{\mu\nu}, \varphi \right)_{\mu\nu}$, the large quantum correction to the vacuum energy is cancelled. After the work of Ref.~\cite{Kaloper:2013zca} appeared, several extensions of the model have been considered \cite{Kaloper:2014dqa,Kaloper:2015jra}. In Ref.~\cite{Kaloper:2015jra}, instead of the global variables $\Lambda$ and $\sigma$, 4-form fields were introduced and the model can be written in a totally local form. Even in the model of Ref.~\cite{Kaloper:2015jra}, the constraint corresponding to (\ref{03}) is global and there is a problem related to the causality. On the other hand, unimodular gravity models, which have properties similar to the sequestering models, have been considered since a long time ago \cite{Anderson:1971pn,Buchmuller:1988wx,Henneaux:1989zc,Unruh:1988in,Ng:1990xz,Finkelstein:2000pg,Alvarez:2005iy,
Alvarez:2006uu,Abbassi:2007bq,Ellis:2010uc,Jain:2012cw,Singh:2012sx,Kluson:2014esa,Padilla:2014yea,Barcelo:2014mua,
Barcelo:2014qva,Burger:2015kie,Alvarez:2015sba,Jain:2012gc,Jain:2011jc,Cho:2014taa,Basak:2015swx,Gao:2014nia,
Eichhorn:2015bna,Saltas:2014cta}.  In the unimodular gravity, the determinant of the metric is constrained to be unity, like in Eq.~(\ref{Uni1}), which we called unimodular constraint. As we demonstrated in the previous section, in the Lagrangian formalism, the constraint can be realized by using the Lagrange multiplier field $\lambda$ as follows:
\be
\label{UniCC0}
S = \int d^4 x \left\{ \sqrt{-g} \left( \mathcal{L}_\mathrm{gravity} - \lambda \right) + \lambda \right\} 
+ S_\mathrm{matter} \, .
\ee
In Eq.~(\ref{UniCC0}), the term $S_\mathrm{matter}$ is the action of matter fluids present, and $\mathcal{L}_\mathrm{gravity}$ denotes all the Lagrangian densities of arbitrary gravity models. Upon variation of action (\ref{UniCC0}) with respect to the Lagrange multiplier $\lambda$, we obtain the unimodular constraint of Eq.~(\ref{Uni1}). We may write the gravity Lagrangian density $\mathcal{L}_\mathrm{gravity}$ in the way that it appears as the sum of the cosmological constant $\Lambda$ and of another part $\mathcal{L}_\mathrm{gravity}^{(0)}$, as follows,
\be
\label{UniCC1}
\mathcal{L}_\mathrm{gravity} = \mathcal{L}_\mathrm{gravity}^{(0)} - \Lambda \, .
\ee
In addition, we cam also redefine the Lagrange multiplier field $\lambda$ to be $\lambda \to \lambda - \Lambda$. Consequently, the action (\ref{UniCC0}) can be rewritten as follows,
\be
\label{UniCC2}
S = \int d^4 x \left\{ \sqrt{-g} \left( \mathcal{L}_\mathrm{gravity}^{(0)} - \lambda \right) + \lambda \right\} 
+ S_\mathrm{matter} + \Lambda \int d^4 x \, .
\ee
Due to the fact that the last term $\Lambda \int d^4 x$ does not depend on any dynamical variable, we may drop the last term. This indicates that the cosmological constant $\Lambda$ does not affect the dynamics. We should note that the cosmological constant may include the large quantum corrections from matter fields to the vacuum energy. Since the cosmological constant $\Lambda$ does not affect the dynamics, the large quantum corrections can be chosen to vanish. For a relevant study to the one presented in this section, see also \cite{Nojiri:2016mlb}

\section{Analysis of the Unimodular $F(R)$ Gravity Formalism Implications}

In the previous sections we described in some detail the unimodular $F(R)$ gravity formalism, and we derived the general equations of motion, however, we did not discuss any fundamental physical implications of the unimodular $F(R)$ formalism. Particularly there are three fundamental questions which should be carefully addressed in the context of every new theoretical framework, with the fist question being related to Newton's law modification. Particularly, the question is how the Newtonian potential is affected by unimodular gravity. Closely related to this question, is the behavior of the graviton field. The second question is related to  the existence of ghost propagating modes, do these exist and if yes, what are the implications of these modes? The third question is related to the matter stability of the theory, and particularly the question is whether the matter stability conditions of the usual $F(R)$ gravity theory suffice to describe the unimodular $F(R)$ gravity. In this section we shall try to address in some detail these questions. 

We start our analysis with the Newton law question. The unimodular constraint (\ref{Uni1}) can be realized by using the Lagrange multiplier field $\lambda$, as we already discussed, and by also taking into account the presence of matter fluids, the total unimodular $F(R)$ gravity with matter fluids can be written, 
\be
\label{UF2}
S = \int d^4 x \left\{ \sqrt{-g} \left( \frac{F(R)}{2\kappa^2} - \lambda \right) + \lambda \right\} 
+ S_\mathrm{matter} \left( g_{\mu\nu}, \Psi \right)\, ,
\ee
where the action $S_\mathrm{matter}$ denotes the action for all matter fluids present, while $\Psi$ expresses the matter fluids. As in the usual $F(R)$ gravity case, we may rewrite the action in the scalar-tensor form as follows,
\be
\label{UF3}
S = \int d^4 x \left\{ \sqrt{-g} \left( \frac{1}{2\kappa^2} \left( R 
- \frac{3}{2} g^{\mu\nu} \partial_\mu \phi \partial_\nu \phi - V(\phi) \right)- \lambda \e^{2\phi} \right) + \lambda \right\} 
+ S_\mathrm{matter} \left( \e^\phi g_{\mu\nu}, \Psi \right)\, ,
\ee
where the potential $V(\phi)$ is equal to,  
\be
\label{UF4}
V(\phi) = \frac{A(\phi)}{F'\left( A \left( \phi \right) \right)} 
 - \frac{F\left( A \left( \phi \right) \right)}{F'\left( A \left( \phi \right) \right)^2}\, ,
\ee
and the function $A(\phi)$ is defined by solving the following algebraic equation, 
\be
\label{UF5}
\phi = - \ln F'(A)\, .
\ee
The unimodular constraint (\ref{Uni1}), which is realized given by the Lagrange multiplier $\lambda$, is now modified to be,
\be
\label{UF6}
\e^{2\phi}\sqrt{-g} = 1 \, .
\ee
Then by deleting the scalar field $\phi$ using (\ref{UF6}), 
the action (\ref{UF3}) can be recast as follows. 
\be
\label{UF7}
S = \int d^4 x \sqrt{-g} \left( \frac{1}{2\kappa^2} \left( R - \frac{3}{32 g^2} 
g^{\mu\nu} \partial_\mu g \partial_\nu g - V\left(\frac{1}{4}\ln \left( - g \right) \right) \right) 
\right) 
+ S_\mathrm{matter} \left( \left( -g \right)^\frac{1}{4} g_{\mu\nu}, \Psi \right)\, .
\ee
At this point it is easy to find the Newtonian limit of the unimodular $F(R)$ scalar-tensor theory, and to this end we consider the following perturbation of the metric $g_{\mu\nu}$ around the background metric $g_{\mu\nu}^{(0)}$,  
\be
\label{UF8}
g_{\mu\nu} = g_{\mu\nu}^{(0)} + h_{\mu\nu}  \, .
\ee
In the small background curvature regime, the background metric can be assumed to be flat, that is, $g_{\mu\nu}^{(0)} = \eta_{\mu\nu}$. By making this assumption, we obtain the following expression of the scalar curvature,
\be
\label{UF10}
\sqrt{- g} R \sim - \frac{1}{2} \partial_\lambda h_{\mu\nu} \partial^\lambda h^{\mu\nu} 
+ \partial_\lambda h^\lambda_{\ \mu} \partial_\nu h^{\mu\nu} 
 - \partial_\mu h^{\mu\nu} \partial_\nu h + \frac{1}{2}\partial_\lambda h \partial^\lambda h  \, ,
\ee
where $h$ denotes the trace of the perturbation $h_{\mu\nu}$, $h \equiv \eta^{\rho\sigma} h_{\rho\sigma}$. By assuming a nearly flat background, we easily find that,
\be
\label{UF10b}
V(0)=V'(0)=0\, ,
\ee
and the potential $V$ is approximately equal to,  
\be
\label{UF10c}
V \sim \frac{1}{2}m^2 h^2\, .
\ee
Then the linearized action has the following form,
\begin{align}
\label{UF12}
S =& \frac{1}{2\kappa^2} \int d^4 x \left\{ 
  - \frac{1}{2} \partial_\lambda h_{\mu\nu} \partial^\lambda h^{\mu\nu} 
+ \partial_\lambda h^\lambda_{\ \mu} \partial_\nu h^{\mu\nu} 
 - \partial_\mu h^{\mu\nu} \partial_\nu h + \frac{1}{2}\partial_\lambda h \partial^\lambda h  
 - \frac{3}{32}\partial_\mu h \partial^\mu h - \frac{1}{2}m^2 h^2 \right\} \nn
& + S_\mathrm{matter} \left( \eta_{\mu\nu} + h_{\mu\nu} - \frac{1}{4}\eta_{\mu\nu} h , \Psi \right)\, .
\end{align}
By varying the action (\ref{UF12}) with respect to the metric perturbation $h_{\mu\nu}$, we obtain the following equations of motion,
\be
\label{UF13}
\partial_\lambda \partial^\lambda h_{\mu\nu} 
 - \partial_\mu \partial^\lambda h_{\lambda\nu}
 - \partial_\nu \partial^\lambda h_{\lambda\mu} 
+ \partial_\mu \partial_\nu h 
+ \eta_{\mu\nu} \partial^\rho \partial^\sigma h_{\rho\sigma} 
 - \frac{13}{16} \eta_{\mu\nu} \partial_\lambda \partial^\lambda h  - m^2 \eta_{\mu\nu} h 
= \kappa^2 \left( T_{\mu\nu} - \frac{1}{4} \eta_{\mu\nu} T \right) \, .
\ee
Note that in the equations of motion (\ref{UF13}), $T_{\mu\nu}$ is the energy-momentum tensor of the matter and $T$ is the trace of 
$T_{\mu\nu}$, with $T \equiv \eta^{\rho\sigma} T_{\rho\sigma}$. Multiplying Eq.~(\ref{UF13}) by $\eta^{\mu\nu}$, we obtain
\be
\label{UF13b}
0 = - \frac{5}{4} \partial_\lambda \partial^\lambda h - 4 m^2 h + 2 \partial^\mu \partial^\nu h_{\mu\nu}\, .
\ee
In order to investigate how Newton's law behaves, we consider the point-like source at the origin, with the components of the corresponding energy-momentum tensor being equal to,
\be
\label{UF14}
T_{00} = M \delta \left( \br \right)\, , \quad T_{ij} = 0 \, \left(i,j=1,2,3\right) \, .
\ee
In the following we consider only static solutions of (\ref{UF13}), with the $(0,0)$, $(i,j)$, and $(0,i)$ components of Eq.~(\ref{UF13}) and Eq.~(\ref{UF13b}) having  the following form:
\begin{align}
\label{UF15}
& \partial_i \partial^i h_{00} - \partial^i \partial^j h_{ij} + \frac{13}{16} \partial_i \partial^i h  
+ m^2 h = \frac{3 \kappa^2}{4} M \delta \left( \br \right) \, , \\
\label{UF16}
& \partial_k \partial^k h_{ij} - \partial_i \partial^k h_{kj} - \partial_j \partial^k h_{ki} 
+ \partial_i \partial_j h + \delta_{ij} \partial^k \partial^l h_{kl} 
 - \frac{13}{16} \delta_{ij} \partial_k \partial^k h  - m^2 \delta_{ij} h 
= \frac{\kappa^2}{4} M \delta \left( \br \right) \, , \\
\label{UF17}
& \partial_j \partial^j h_{0i} - \partial_i \partial^k h_{k0} = 0 \, ,\\
\label{UF18}
& - \frac{5}{4} \partial_k \partial^k h - 4 m^2 h + 2 \partial^i \partial^j h_{ij} = 0 \, .
\end{align}
In the usual Einstein-Hilbert gravity, there exist four gauge degrees of the freedom, but in the case of unimodular $F(R)$ gravity, there only three gauge degrees of freedom exist, due to the unimodular constraint (\ref{Uni1}). Then we now impose the following three gauge conditions, 
\be
\label{UF19}
\partial^i h_{ij} = 0 \, .
\ee
In effect, Eq.~(\ref{UF18}) becomes equal to, 
\be
\label{UF20}
 - \frac{5}{4} \partial_k \partial^k h - 4 m^2 h = 0 \, ,
\ee
and by using an appropriate boundary condition we obtain,
\be
\label{UF21}
h=0 \, .
\ee
Therefore, by using Eqs. (\ref{UF19}) and (\ref{UF21}), we can rewrite Eqs.~(\ref{UF15}), (\ref{UF16}), 
and (\ref{UF17}) in the following way, 
\begin{align}
\label{UF22}
& \partial_i \partial^i h_{00} = \frac{3 \kappa^2}{4} M \delta \left( \br \right) \, , \\
\label{UF23}
& \partial_k \partial^k h_{ij} = \frac{\kappa^2}{4} M \delta \left( \br \right) \, , \\
\label{UF24}
& \partial_j \partial^j h_{0i} - \partial_i \partial^k h_{k0} = 0 \, .
\end{align}
By also using an appropriate boundary condition, the above equations combined with Eq. (\ref{UF21}) yield, 
\be
\label{UF25}
h_{0i}=0\, , \quad h_{ij} = \frac{1}{3} \delta_{ij} h_{00}\, .
\ee
We define the Newtonian gravitational potential $\Phi$ by $h_{00} = 2 \Phi$, so Eq.~(\ref{UF22}) gives the Poisson equation for 
the Newtonian potential $\Phi$, that is,
\be
\label{UF26}
\partial_i \partial^i \Phi = \frac{3 \kappa^2}{8} M \delta \left( \br \right) \, .
\ee
Then, if we redefine the gravitational constant $\kappa$ by 
\be
\label{UF27}
\frac{3 \kappa^2}{4} \to \kappa^2 = 8\pi G \, ,
\ee
we obtain the standard Poisson equation for the Newtonian potential $U$,
\be
\label{UF28}
\partial_i \partial^i \Phi = 4\pi G M \delta \left( \br \right) \, ,
\ee
whose solution is given by, 
\be
\label{UF29}
\Phi = - \frac{GM}{r}\, .
\ee
Hence what we actually demonstrated is that the Newtonian limit of the unimodular $F(R)$ gravity is quite different from the usual $F(R)$ gravity, in which case, the propagation of the scalar mode $\phi$ in (\ref{UF5}) gives a correction to Newton's law. In the case of unimodular $F(R)$ gravity, because the unimodular condition (\ref{Uni1}) can 
be rewritten as in Eq. (\ref{UF6}), the degree of the freedom of the scalar mode $\phi$ is eliminated, and the scalar mode $\phi$ does not propagate. In effect, there is no correction to the Newtonian potential, so this is the first difference between unimodular and ordinary $F(R)$ gravity. 

Having discussed the Newtonian approximation and the corrections to Newton's law, now what remains is to discuss the propagating modes in the unimodular $F(R)$ gravity, focusing to the graviton and also to ghost modes (if any). We shall address this question by assuming that no matter fluids are present. Then, by putting $T_{\mu\nu}=0$ in (\ref{UF13}), we obtain the following linearized equation, 
\be
\label{UF13V}
0=\partial_\lambda \partial^\lambda h_{\mu\nu} 
 - \partial_\mu \partial^\lambda h_{\lambda\nu}
 - \partial_\nu \partial^\lambda h_{\lambda\mu} 
+ \partial_\mu \partial_\nu h 
+ \eta_{\mu\nu} \partial^\rho \partial^\sigma h_{\rho\sigma} 
 - \frac{13}{16} \eta_{\mu\nu} \partial_\lambda \partial^\lambda h  - m^2 \eta_{\mu\nu} h \, .
\ee
Multiplying Eq.~(\ref{UF13V}) by $\eta^{\mu\nu}$, we obtain,
\be
\label{UF13b}
0 = - \frac{5}{4} \partial_\lambda \partial^\lambda h - 4 m^2 h + 2 \partial^\mu \partial^\nu h_{\mu\nu}\, .
\ee
We also impose the gauge condition (\ref{UF19}) and we decompose the metric $h_{\mu\nu}$ in the following way,
\be
\label{UFRP1}
h_{ij} = \hat h_{ij} + \frac{1}{3}\delta_{ij} A 
+ \partial_i B_j + \partial_j B_i 
+ \partial_i \partial_j C - \frac{1}{3} \delta_{ij} \partial_k \partial^k C \, , \quad 
h_{0i} = D_i + \partial_i E\, , \quad h_{00} = F \, ,
\ee
where the tensor $\hat h_{ij}$ describes the massless graviton and $\hat H_{ij}$, $B_i$, and $D_i$ satisfy the 
following conditions, 
\be
\label{UFRP2}
\partial^i \hat h_{ij} = 0 \, , \quad \hat h \equiv \hat h^i_{\ i}=0\, , \quad 
\partial^i B_i = \partial^i D_i = 0 \, .
\ee
Then, the gauge condition (\ref{UF19}) takes the following form,
\be
\label{UFRP3}
0 = \frac{1}{3} \partial_j A + \partial_i \partial^i B_j + \frac{2}{3} \partial_j \partial_i \partial^i C \, ,
\ee
and by differentiating we obtain, 
\be
\label{UFRP3b}
0 = \partial_j \partial^j \left( \frac{1}{3} A + \frac{2}{3} \partial_i \partial^i C \right) \, .
\ee
In effect, by using an appropriate boundary condition, we find,
\be
\label{UFRP4}
0 = \frac{1}{3} A + \frac{2}{3} \partial_i \partial^i C \, ,
\ee
and by substituting (\ref{UFRP4}) into (\ref{UFRP3}), we obtain,
\be
\label{UFRP5}
0 = \partial_i \partial^i B_j \, .
\ee
By using a proper boundary condition, we find get,
\be
\label{UFRP6}
B_j = 0 \, ,
\ee
and consequently, the graviton tensor field $\hat h_{ij}$ receives the following form,
\be
\label{UFRP7}
h_{ij} = \hat h_{ij} + \partial_i \partial_j C -  \delta_{ij} \partial_k \partial^k C \, .
\ee
and therefore the following holds true, 
\be
\label{UFRP18b}
0 = \left( \partial_0^2 + \partial_j \partial^j  \right) C \, ,
\ee
We should note the $+$ sign in Eq. (\ref{UFRP18b}), indicates that the mode $C$ does not propagate, but also it indicates that a growing unstable mode exists. Then, we impose by hand the following constraint, 
\be
\label{UFRP19}
C = 0 \, ,
\ee
and also, Eq.~(\ref{UFRP1}) also indicates that, 
\be
\label{UFRP20}
E=0 \, .
\ee
By substituting (\ref{UFRP19}) into (\ref{UFRP2}), we obtain,
\be
\label{UFRP21}
0 = \left( \partial_\lambda \partial^\lambda - \frac{16}{3} m^2 \right) F\, .
\ee
Therefore, $F$ seems to be the propagating scalar mode with mass $m_F^2 =  \frac{16}{3} m^2$. 
By using Eqs.~(\ref{UFRP2}), (\ref{UFRP18b}), (\ref{UFRP19}), (\ref{UFRP20}), and (\ref{UFRP21}) and multiplying the resulting equation by $\delta^{ij}$, we finally find, 
\be
\label{UFRP22}
0=\partial_k \partial^k F \, ,
\ee
and therefore $F$ does not correspond to the propagating mode but we get, 
\be
\label{UFRP23}
F=0\, .
\ee
and consequently we find, 
\be
\label{UFRP22}
0=\partial_\lambda \partial^\lambda \hat h_{ij} \, ,
\ee
which tells that the tensor field $\hat h_{ij}$ describes for sure a massless graviton, and it is the only propagating mode. In effect, this means that only the graviton corresponds to a propagating mode and also no propagating ghost degree of freedom exists. In addition, the only source of instability in the system is the mode $C$.

Finally, let us discuss in brief the matter stability issue, which for the case of ordinary $F(R)$ gravity it is discussed in Refs. \cite{Nojiri:2010wj,Nojiri:2013zza}. By varying the unimodular $F(R)$ gravity action with matter fluids appearing in Eq. (\ref{UF2}) with respect to the metric $g_{\mu \nu}$, we obtain the following equations of motion,
\be
\label{Uni12}
0=\frac{1}{2}g_{\mu\nu} \left( F(R) - \lambda \right) - R_{\mu\nu} F'(R)
+ \nabla_\mu \nabla_\nu F'(R) - g_{\mu\nu}\nabla^2 F'(R) 
+ \frac{\kappa^2}{2} T_{\mu\nu}
\, .
\ee
By multiplying (\ref{Uni12}) with $g^{\mu\nu}$, we obtain,
\be
\label{MI1}
0= 2 \left( F(R) - \lambda \right) - R F'(R) - 3 \nabla^2 F'(R) 
+ \frac{\kappa^2}{2} \, .
\ee
In case of the ordinary $F(R)$ gravity, which corresponds to the case of $\lambda=0$, 
it has been shown in the literature \cite{Nojiri:2010wj,Nojiri:2013zza,Dolgov:2003px}, that if the matter density is large enough, as on the earth, a strong instability occurs. However, in the case of unimodular $F(R)$ gravity, the role of Eq.~(\ref{MI1}) is to determine $\lambda$. Therefore, by using (\ref{Uni12}) and (\ref{MI1}), and by eliminating $\lambda$, we obtain,
\be
\label{MI2}
0= - R_{\mu\nu} F'(R) + \nabla_\mu \nabla_\nu F'(R) - \frac{1}{4} g_{\mu\nu}
\left( R F'(R) - \nabla^2 F'(R)  \right) 
+ \frac{\kappa^2}{2} \left( T_{\mu\nu} - \frac{1}{4}T g_{\mu\mu} T \right) \, .
\ee
The trace of Eq.~(\ref{MI2}) trivially vanishes and therefore in order to discuss about the matter instability issue, we need another formalism, since the standard approach of the ordinary $F(R)$ gravity does not suffice. Hence the matter stability problem in unimodular $F(R)$ gravity cannot be done by using the usual ordinary $F(R)$ gravity techniques, and a new approach is needed. This, however, exceeds the purposes of this paper, since this is a completely new study which should be done separately.

\section{General Features of Bouncing Cosmologies and Realization from Unimodular $F(R)$ Gravity}

As we previously discussed, the bouncing cosmologies stand as a quite appealing alternative scenario to the standard inflationary scenario \cite{Mukhanov:2005sc,Gorbunov:2011zzc,Linde:2014nna,Lyth:1998xn,Bamba:2015uma}, with the attractive feature of not having an initial singularity, being the most appealing feature of the big bounce scenario. We use four paradigms of bouncing cosmologies, and specifically, the matter bounce scenario \cite{Brandenberger:2012zb,Quintin:2014oea,Cai:2011ci,Cai:2011zx,Bamba:2012ka,deHaro:2012xj,deHaro:2014jva,
deHaro:2015wda,Bamba:2013fha,Barragan:2009sq,
Brandenberger:2012zb}, the singular bounce \cite{Odintsov:2015ynk,Oikonomou:2015qfh,Oikonomou:2015qha,Odintsov:2015zza} and also the superbounce scenario \cite{Koehn:2013upa} and the symmetric bounce. For these we discuss in detail how the Hubble horizon evolves during the bounce evolution and we also discuss the cosmological perturbations issue \cite{Mukhanov:1990me,Brandenberger:1983tg,Brandenberger:1983vj,Baumann:2009ds}, in order to extract enough information on how to connect early-time phenomena with present time observations. But the bouncing cosmologies in general should also address consistently all the theoretical problems that the standard inflationary scenario solved in the first place. It is worth to discuss these issues in the context of the standard inflationary paradigm and we immediately compare the bouncing cosmologies with the inflationary picture \cite{Mukhanov:2005sc,Gorbunov:2011zzc,Linde:2014nna,Lyth:1998xn,Bamba:2015uma,Mukhanov:1990me,Brandenberger:1983tg,Brandenberger:1983vj,Baumann:2009ds}. For reviews and important papers on cosmological perturbations we refer the reader to \cite{Mukhanov:2005sc,Gorbunov:2011zzc,Linde:2014nna,Lyth:1998xn,Bamba:2015uma,Mukhanov:1990me,Brandenberger:1983tg,Brandenberger:1983vj,Baumann:2009ds}, and for an insightful recent study that it is worth mentioning, see \cite{Wetterich:2015iea}.

Our ability to make predictions on early-time physics is owing to the linear cosmological perturbation theory \cite{Mukhanov:1990me,Brandenberger:1983tg,Brandenberger:1983vj,Baumann:2009ds}, and in the context of the inflationary paradigm, the primordial quantum fluctuations of the comoving curvature were at subhorizon scales during inflation. It is exactly these fluctuations which are relevant for present time observations, and during inflation, the wavelength of these fluctuations were at subhorizon scales, meaning that it was much more smaller in comparison to the Hubble radius, which is defined to be $R_H=\frac{1}{a(t)H(t)}$, with $a(t)$ and $H(t)$ being the scale factor and Hubble rate respectively. Quantitatively, before the inflationary era and after the initial singularity, the Hubble radius was very large and the primordial quantum fluctuating modes were at subhorizon scales, since the corresponding comoving wave number satisfied,
\begin{equation}\label{subhorizonsscalaesk}
k\gg H(t)a(t)\, .
\end{equation}
or equivalently, $\lambda \ll \left( H(t)a(t) \right)^{-1}$, with $\lambda$ being the corresponding wavelength. During the inflationary era, the Hubble radius decreased very much, so eventually the primordial modes exit the horizon when $k=a(t_H)H(t_H)$. After the horizon crossing, the primordial modes become superhorizon modes and will satisfy
\begin{equation}\label{superhorizon1}
k\ll a(t)H(t)\, ,
\end{equation}
until the horizon re-entry. All the aforementioned features are properly addressed by the inflationary paradigm, so we may conclude that the vital features of a viable cosmological evolution are two: First, the Hubble horizon should contract at early times and second, the Hubble horizon should eventually re-expand. Let us now discuss these issues in the context of the bouncing cosmologies.

The matter bounce cosmological scenario \cite{Brandenberger:2012zb,Quintin:2014oea,Cai:2011ci,Cai:2011zx,Bamba:2012ka,deHaro:2012xj,deHaro:2014jva,
deHaro:2015wda,Bamba:2013fha,Barragan:2009sq,
Brandenberger:2012zb} results from Loop Quantum Cosmology (LQC) \cite{Ashtekar:2011ni,Ashtekar:2007tv,Ashtekar:2011ni,Corichi:2009pp,Singh:2009mz,Bojowald:2008ik}, and is a viable alternative scenario to inflation, with the interesting feature of compatibility with the Planck \cite{Ade:2015lrj,Planck:2013jfk} observational data. The scale factor and the corresponding Hubble rate of the matter bounce scenario are,
\begin{equation}
\label{matterbouncescale}
a(t)=\left(\frac{3}{2} \rho_c t^2+1\right)^{\frac{1}{3}}\, , \quad H(t)=\frac{2 t \rho_c}{2+3 t^2 \rho_c}\, ,
\end{equation}
with $\rho_c$ being a critical energy density determined by the LQC underlying theory.
\begin{figure}[h] \centering
\includegraphics[width=12pc]{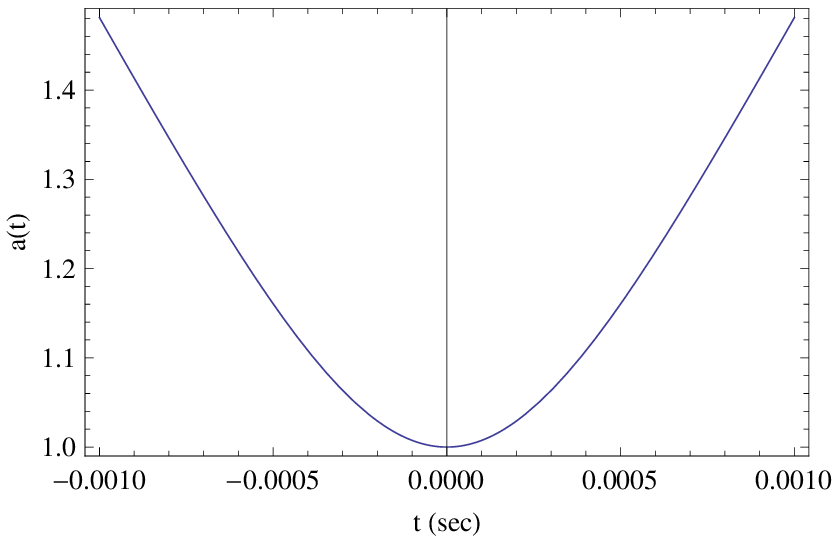}
\includegraphics[width=12pc]{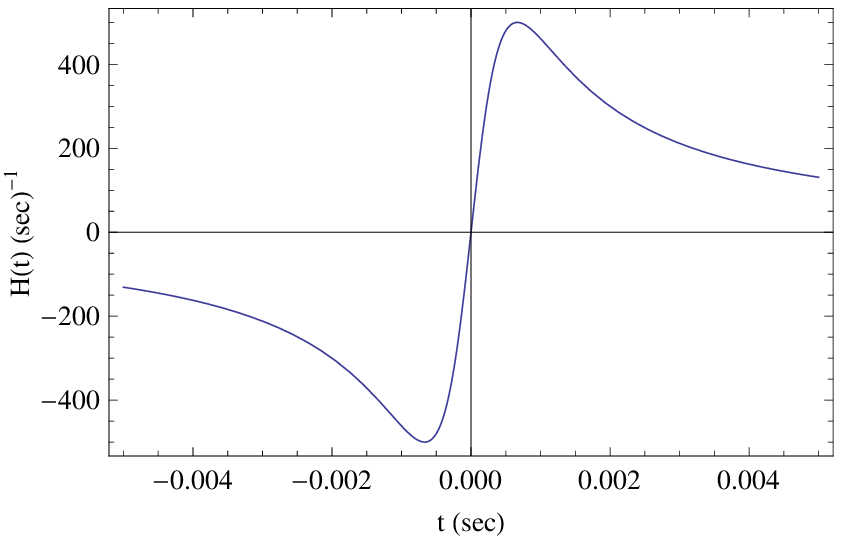}
\includegraphics[width=12pc]{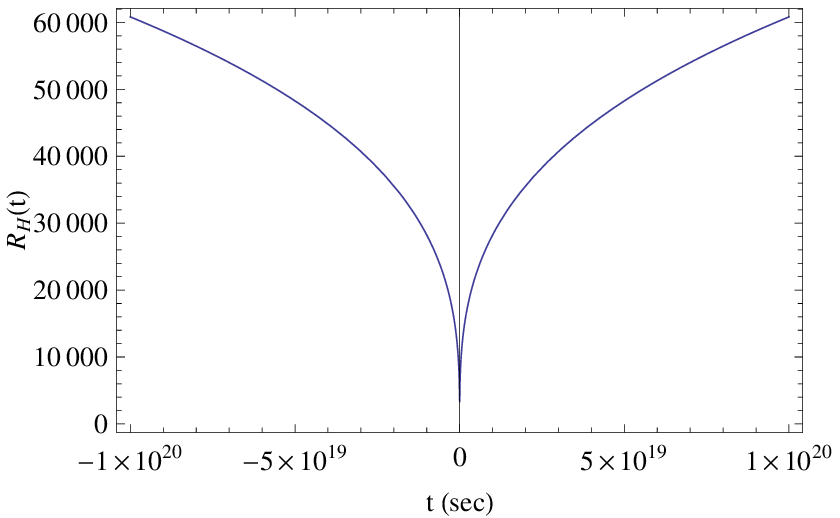}
\caption{The scale factor $a(t)$ (left plot), the Hubble rate
(middle plot) and the Hubble radius $R_H(t)$ (right plot) as functions of the cosmological time $t$, for
the matter bounce scenario $a(t)=\left(\frac{3}{2} \rho_c t^2+1\right)^{\frac{1}{3}}$.}
\label{mattbounce1}
\end{figure}
In Fig.~\ref{mattbounce1}, we have plotted the scale factor, the Hubble rate and the Hubble radius, as functions of the cosmological time, and it can be seen, the bounce cosmology conditions are satisfied, and also the scale factor decreases for $t<0$ and increases for $t>0$. Let us study now discuss the evolution of the Hubble horizon, which determines the cosmological perturbation behavior and especially determines the exact time that corresponds to the generation of the cosmological perturbations. In Fig.~\ref{mattbounce1}, right plot, we plotted the Hubble radius $R_H(t)$ as a function of the cosmological time $t$, for $\rho_c=10^6\, \mathrm{sec}^{-2}$, and as it can be seen, before the bouncing point, which is at $t=0$, the Hubble radius starts from an infinite size at $t\rightarrow -\infty$, and gradually decreases until the bouncing point and after the bounce the Hubble radius increases again. 

The realization of the matter bounce scenario from unimodular $F(R)$ gravity is straightforward and can be done in two limiting cases, for times near the bounce and for large cosmic times values. In Table \ref{newtable} later on in this section, we have gathered the results for the $F(R)$ gravity, and the Lagrange multiplier can easily be found by using the formalism we presented earlier. The resulting form of the unimodular $F(R)$ gravity in the limit $t\to 0$, is a complex function of the Ricci scalar. Let us recall from the existing literature what does a complex $F(R)$ gravity implies. As was demonstrated in Ref.~\cite{Briscese:2006xu}, a complex $F(R)$ gravity is related to a phantom scalar-tensor theory, since these two theories are mathematically equivalent. Also as was proved in the same work \cite{Briscese:2006xu}, even in the case that the scalar potential is even, and the corresponding $F(R)$ gravity is real, the region where the original scalar tensor theory develops a Big Rip singularity, corresponds to a complex $F(R)$ gravity. Finally, as it can be seen in Table \ref{newtable}, by comparing the resulting forms of the unimodular $F(R)$ gravity for both the limits $t\to 0$ and $t\to \infty$, these are different from the corresponding ordinary $F(R)$ gravity which was found in Ref.~\cite{Odintsov:2014gea}.

Another scenario with interesting phenomenology is the superbounce scenario \cite{Koehn:2013upa,Odintsov:2015uca,Oikonomou:2014yua}, which was firstly studied in the context of some ekpyrotic scenarios \cite{Koehn:2013upa}. The scale factor and the Hubble rate for the superbounce are given below,
\begin{equation}
\label{superbouncescale}
a(t)=(-t+t_s)^{\frac{2}{c^2}}\, , \quad H(t)=-\frac{2}{c^2 (-t+t_s)}\, ,
\end{equation}
with $c$ being an arbitrary parameter of the theory while the bounce in this case occurs at $t=t_s$.
\begin{figure}[h] \centering
\includegraphics[width=12pc]{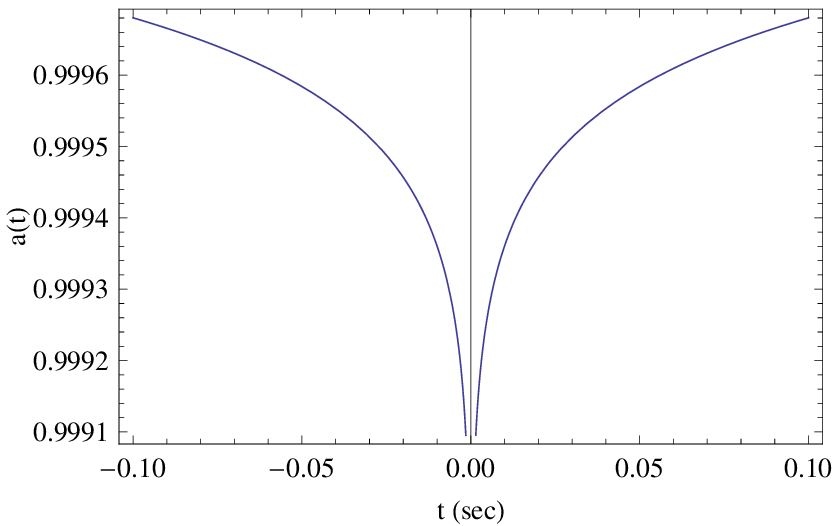}
\includegraphics[width=12pc]{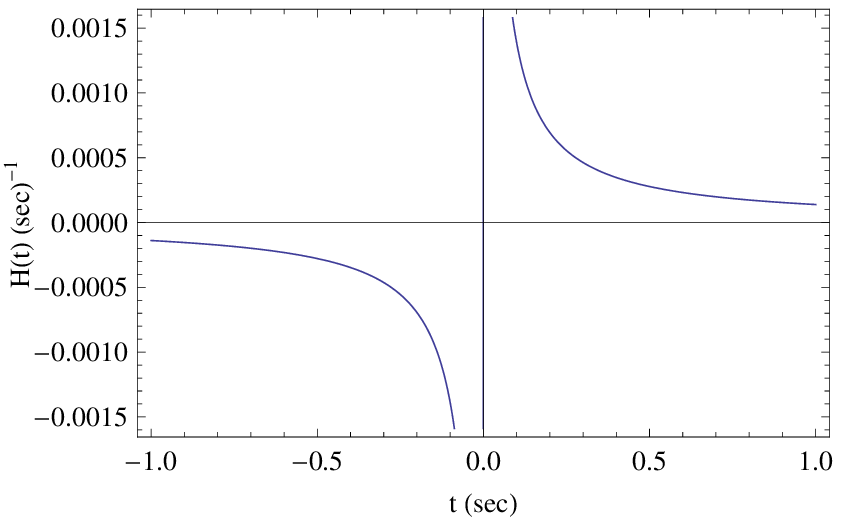}
\includegraphics[width=12pc]{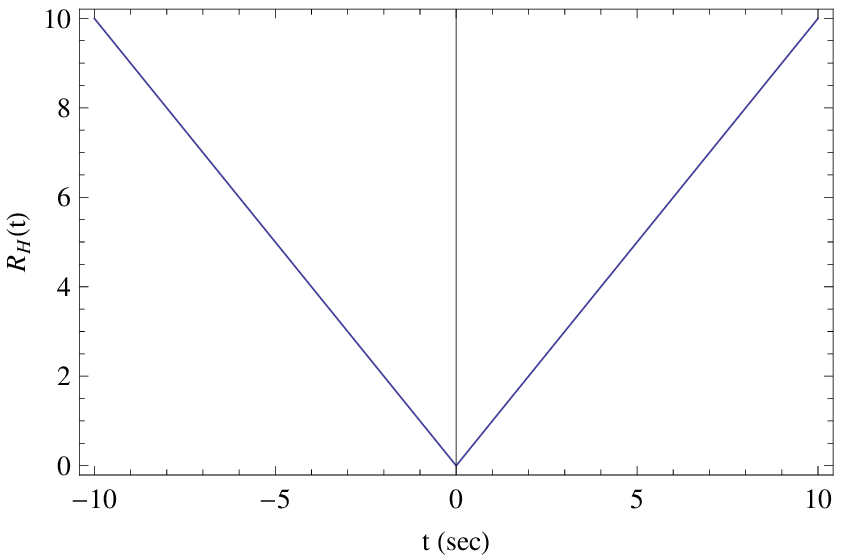}
\caption{The scale factor $a(t)$ (left plot), the Hubble rate (middle plot), and the Hubble radius $R_H(t)$ (right plot) as functions of the cosmological time $t$ for
the superbounce scenario $a(t)=(-t+t_s)^{\frac{2}{c^2}}$.}
\label{superbounce1}
\end{figure}
In Fig.~\ref{superbounce1}, we have plotted the time dependence of the scale factor, the Hubble rate and the Hubble radius, for the superbounce case. It can be seen that in this case too, the bounce cosmology conditions are satisfied, and in addition, the scale factor decreases for $t<0$ and increases for $t>0$, as in every bounce cosmology, so contraction and expansion occurs. In addition, the physics of the cosmological perturbations are the same to the matter bounce case, since the Hubble radius decreases for $t<0$ and increases for $t>0$, so the correct description for the superbounce is the following: Initially, the Universe starts with an infinite Hubble radius, at $t\rightarrow -\infty$, so the primordial modes are at subhorizon scales at that time. Gradually, the Hubble horizon decreases and consequently the modes exit the horizon and possibly freeze. Eventually, after the bouncing point, the Hubble horizon increases again, so it is possible for the primordial modes to reenter the horizon. Hence this model can harbor a conceptually complete phenomenology. The behavior of the Hubble horizon as a function of the cosmological time can be found in Fig.~\ref{superbounce1}, right plot.

In Table \ref{newtable} we present the unimodular $F(R)$ gravity which can realize the superbounce scenario, and the unimodular Lagrange multiplier can easily be found in a similar way. As it can be verified, the resulting expression for the unimodular $F(R)$ gravity is different from the ordinary $F(R)$ result found in Refs.~\cite{Odintsov:2015uca,Oikonomou:2014yua}.

\begin{figure}[h] \centering
\includegraphics[width=12pc]{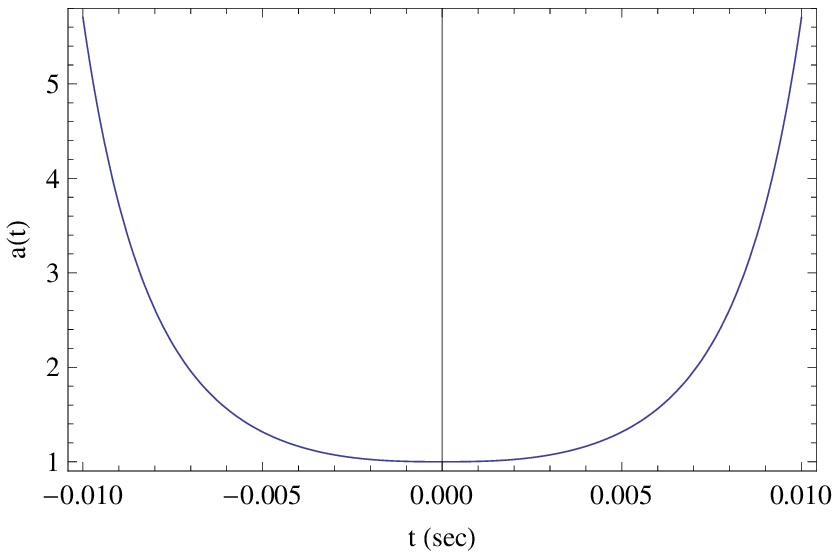}
\includegraphics[width=12pc]{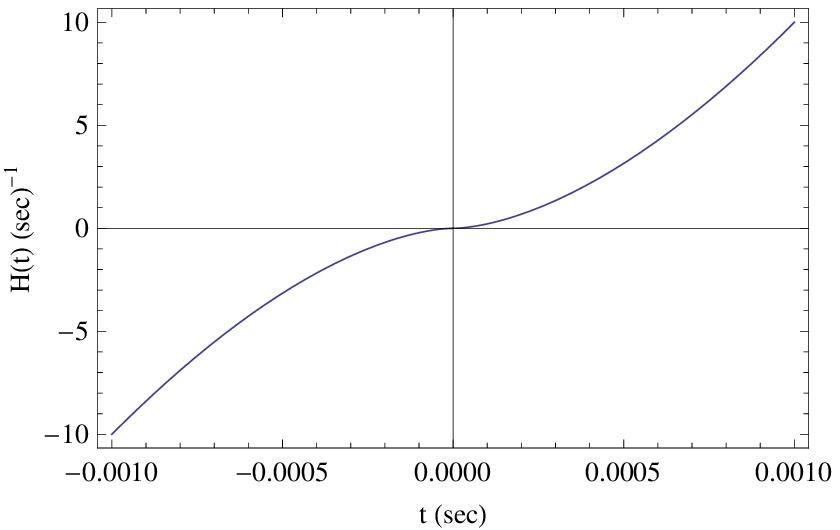}
\includegraphics[width=12pc]{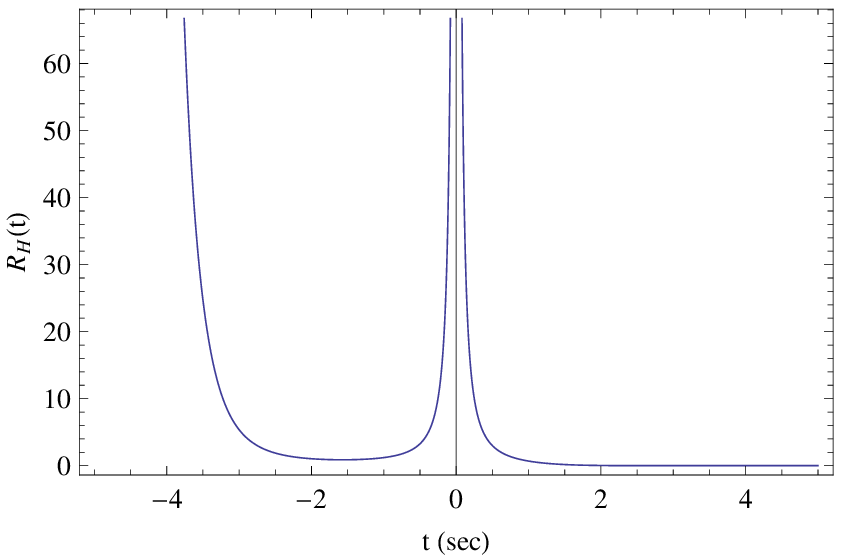}
\caption{The scale factor $a(t)$ (left plot), the Hubble rate
(middle plot) and the Hubble radius $R_H(t)$ (right plot), as functions of the cosmological time $t$, for
the singular bounce scenario $a(t)=\e^{\frac{f_0}{\alpha +1} (t-t_s)^{\alpha +1 }}$.}
\label{singbounce1}
\end{figure}

The singular bounce is a peculiar case of a bounce as it proves, which was extensively studied in a recent series of papers \cite{Odintsov:2015ynk,Oikonomou:2015qfh,Oikonomou:2015qha,Odintsov:2015zza}. This bounce cosmology avoids the initial singularity, which is a crushing singularity, but a Type IV singularity occurs at the bouncing point, see \cite{Odintsov:2015ynk,Oikonomou:2015qfh,Oikonomou:2015qha,Odintsov:2015zza} for details. In this case, the phenomenology is rich, as was demonstrated in \cite{Odintsov:2015ynk,Oikonomou:2015qfh,Oikonomou:2015qha,Odintsov:2015zza}, and the cosmological scenarios can vary. As was proven in \cite{Odintsov:2015ynk,Oikonomou:2015qfh,Oikonomou:2015qha,Odintsov:2015zza}, the singular bounce leads to a non-scale invariant spectrum, if the perturbations originate near the bouncing point, so the singular bounce scenario has to be combined with another scenario in order it leads to a viable cosmology. But let us see in detail the behavior of the bounce, since another interesting scenario is revealed from this study, which however we shall develop in detail in a future work. But in order to reveal this alternative scenario, let us recall the essential information of the singular bounce. The singular bounce scale factor and Hubble rate are equal to,
\begin{equation}
\label{singularbscale}
a(t)=\e^{\frac{f_0}{\alpha +1} (t-t_s)^{\alpha +1 }} \, ,\quad H(t)=f_0\left( t-t_s\right)^{\alpha }\, ,
\end{equation}
with $f_0$ an arbitrary positive real number, and $t_s$ is the time instance at which the bounce occurs and also coincides with the time that the singularity occurs. In order for a Type IV singularity to occur, the parameter $\alpha$ has to satisfy $\alpha>1$. In addition, in order for the singular bounce to obey the bounce cosmology conditions, the parameter $\alpha$ has to be chosen in the following way,
\begin{equation}
\label{alpahhacond}
\alpha=\frac{2n+1}{2m+1}\, ,
\end{equation}
with $n$ and $m$ integers chosen so that $\alpha>1$. For example, for $\alpha=\frac{5}{3}$, the time dependence of the scale factor, the Hubble rate and of the Hubble radius, are given in Fig.~\ref{singbounce1}, and as it can be seen, the bounce conditions are satisfied, and in this case, contraction and expansion occurs.
The singular bounce however, in contrast to the previous two cases, generates a peculiar Hubble radius behavior. In order to make this clear, in Fig.~\ref{singbounce1} right plot, we plotted the Hubble radius as a function of time. As it can be seen, the behavior of the Hubble radius is different in comparison to the previous two cases. Particularly, at $t\rightarrow -\infty$, the Hubble radius is infinite, and gradually decreases until a minimal size, but near the bouncing point it increases and blows up at exactly the bouncing point. Eventually, after the bouncing point it decreases gradually. This is different in comparison to other bouncing cosmologies, and this can be seen by comparing directly the behavior of the Hubble radius.

In Table \ref{newtable} we present the unimodular $F(R)$ gravity that produces the singular bounce near the bouncing point. As it can be seen from Table \ref{newtable} the resulting $F(R)$ gravity is not the same in comparison to the usual non-unimodular $F(R)$ gravity generating the singular bounce, see for example Refs.~\cite{Odintsov:2015ynk,Oikonomou:2015qfh,Oikonomou:2015qha,Odintsov:2015zza}.

\begin{figure}[h] \centering
\includegraphics[width=12pc]{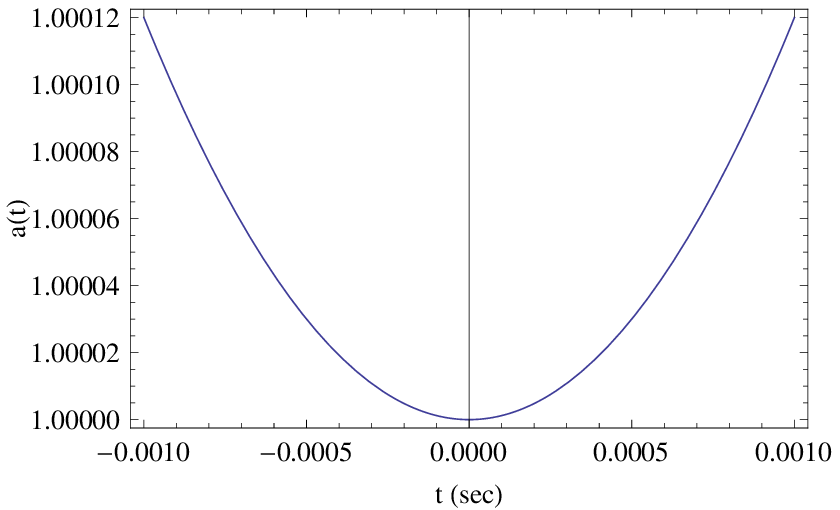}
\includegraphics[width=12pc]{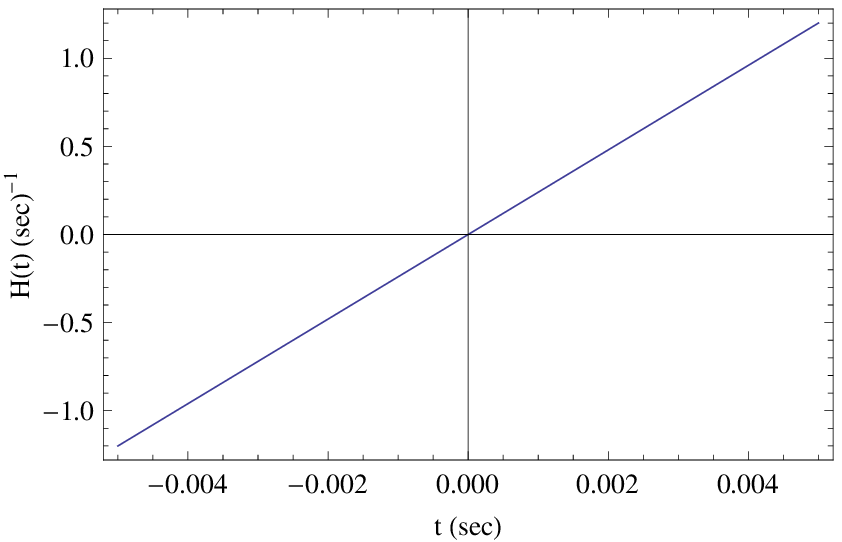}
\includegraphics[width=12pc]{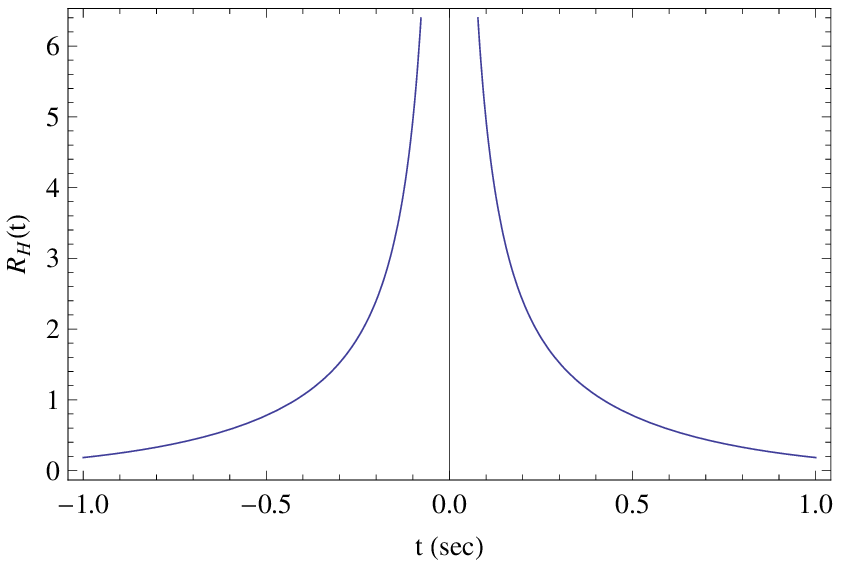}
\caption{The scale factor $a(t)$ (left plot) and the Hubble rate
(middle plot) and the Hubble radius $R_H(t)$ (right plot) for
the symmetric bounce scenario $a(t)=\e^{f_0t^2}$,  as functions of the cosmological time $t$.}
\label{symbounce1}
\end{figure}

The symmetric bounce case is another interesting and simple bouncing cosmology, which was studied in the context of modified gravity in Ref \cite{Bamba:2013fha}. In this case, the scale factor and the Hubble rate are equal to,
\begin{equation}
\label{symmbouncescale}
a(t)=\e^{f_0t^2} \, ,\quad H(t)=2f_0t\, ,
\end{equation}
with the parameter $f_0$ being a real positive number. In Fig.~\ref{symbounce1} we plotted the scale factor, the Hubble rate and the Hubble radius, for the symmetric bounce case, and it can be seen that the bounce cosmology conditions are satisfied and also that contraction and expansion around the bouncing point $t=0$ occurs.
However, in this case the Hubble horizon evolves in a peculiar way as it can be seen in Fig.~\ref{symbounce1}, right plot. Particularly, the Hubble horizon is almost zero at $t\rightarrow -\infty$ and as $t\rightarrow 0$, the Hubble horizon rapidly grows and it blows up at $t=0$. After the bouncing point, the Hubble radius decreases rapidly again, and continues to decrease. It is obvious that this scenario is physically incomplete for two reasons. Firstly, the Hubble horizon from $t\rightarrow -\infty$ never decreases, but increases, so the primordial modes cannot be considered that originate from a time before the bounce. Then the only possibility is that the primordial modes correspond to a time near the bouncing point. Secondly, even in this case, the horizon after $t>0$ decreases, so there is no possibility for horizon reentry of the modes. This means that, this cosmological bounce model should be combined with another cosmological scenario, as was performed for example in Ref.~\cite{Cai:2014jla}. However, in this paper our focus is not to exactly study the bounces, but to see how the bounces can be realized in the context of unimodular $F(R)$ gravity, so for completeness we shall be interested in the realization of the symmetric bounce of Eq.~(\ref{symmbouncescale}).

In Table \ref{newtable} we present the unimodular $F(R)$ gravity which can realize the symmetric bounce scenario near the bouncing point. As it can be seen, the unimodular $F(R)$ gravity is a complex function so the same study we discussed in the matter bounce case, about having complex $F(R)$ gravity, should be done in this case too (see Ref.~\cite{Briscese:2006xu} for details). 
\begin{table*}[h]
    \small
\begin{tabular}{@{}|c|r|rrrrrrrrrr@{}}
        \tableline
        \tableline
        \tableline
        Bounce Type & Form of the unimodular $F(R)$ gravity 
        \\\tableline
        Matter Bounce for $t\to 0$ &  $F(R)=\mathcal{A}_3 R+\frac{4 \mathcal{A}_4 (R-3 \rho_c)^{3/2}}{3 \sqrt{39} \rho_c}$
        \\\tableline
        Matter Bounce for $t\to \infty$ &  $F(R)=\mathcal{A}_1 R^{-\frac{3}{2}\left(\frac{23}{18}-\frac{\sqrt{73}}{18}\right)}+\mathcal{A}_2 R^{-\frac{3}{2}\left(\frac{23}{18}+\frac{\sqrt{73}}{18}\right)}$
        \\\tableline
        Superbounce  &   $F(R) = \Omega_1 \,R^{\frac{3}{4}-\frac{1}{2 c^2}+\frac{\sqrt{4+20 c^2+c^4}}{4 c^2}} +\Omega_2\, R^{\frac{3}{4}-\frac{1}{2 c^2}-\frac{\sqrt{4+20 c^2+c^4}}{4 c^2}}$  
        \\\tableline
        Singular Bounce for $t\to 0$ & $
F'(R)\simeq \left(C_2 -\frac{2^{-\frac{1}{1+\alpha }} C_1 \left(-\frac{f_0}{1+\alpha }\right)^{-\frac{1}{1+\alpha }} \,\, \Gamma\left(\frac{1}{1+\alpha },0\right)}{1+\alpha }\right)\e^{-\frac{2 f_0 \left(6^{-\frac{1}{-1+\alpha }} (1+\alpha )\left(\frac{1}{f_0 \alpha }\right)^{\frac{1}{-1+\alpha }}\right)^{1+\alpha }R^{\frac{1}{-1+\alpha }}}{1+\alpha }} 
 $
        \\\tableline
        \tableline
        Symmetric Bounce for $t\to \infty$ &  $F'(R)=C_2 R+\frac{2^{-\frac{1}{10} i \left(-5 i+\sqrt{5}\right)} C_1 \sqrt{\pi } R}{\,\, \Gamma\left(\frac{3}{4}+\frac{i}{4 \sqrt{5}}\right)}+\frac{\left(\frac{1}{15}-\frac{i}{15}\right) 2^{-2-\frac{i}{2 \sqrt{5}}} \left(-5 i+\sqrt{5}\right) C_1 \sqrt{\frac{\pi }{3}} (-12 f_0+R)^{3/2}}{5^{1/4} \sqrt{f_0} \,\, \Gamma\left(\frac{5}{4}+\frac{i}{4 \sqrt{5}}\right)}$
        \\\tableline
        \tableline
    \end{tabular}
    \caption{\label{newtable} The unimodular $F(R)$ gravities realizing the various bouncing cosmologies. The various parameters can be found in the Appendix.}
\end{table*}

Finally a remark is in order. A question that can be naturally asked is whether there exist a mapping enabling to relate the models studied in unimodular and ordinary $F(R)$ gravity frameworks. The answer seems to be no, since no obvious transformation exists between the two frameworks, and it happens in some cases that the resulting ordinary $F(R)$ and unimodular $F(R)$ descriptions is identical. This behavior happens when the $t-\tau$ relation is trivial, that is when $t\sim \tau+\mathrm{constants}$. In general however, at least to our knowledge, there is no direct correspondence between the two frameworks via a transformation. Indeed, the fact that these theories are just two different models is most easily seen from scalar-tensor presentation which has totally different structure in both cases.

\section{Conclusions}

In this article we investigated how several quite well known bouncing cosmologies can be realized in the context of the unimodular $F(R)$ gravity. After presenting in brief the unimodular $F(R)$ gravity formalism \cite{Nojiri:2015sfd}, we investigated how the following bouncing cosmologies can be realized: the matter bounce scenario \cite{Brandenberger:2012zb,Quintin:2014oea,Cai:2011ci,Cai:2011zx,Bamba:2012ka,deHaro:2012xj,deHaro:2014jva,deHaro:2015wda,Bamba:2013fha,Barragan:2009sq}, the superbounce scenario \cite{Koehn:2013upa}, the singular bounce \cite{Odintsov:2015ynk,Oikonomou:2015qfh,Oikonomou:2015qha,Odintsov:2015zza} and the symmetric bounce \cite{Bamba:2013fha}. For all these cosmologies, we examined the behavior of the Hubble radius, also known as the Hubble horizon, in order to reveal the cosmological era at which the cosmological perturbations are generated. Also we investigated how to realize the aforementioned bouncing cosmologies with unimodular $F(R)$ gravity.

In addition, we discussed how Newton's law is modified in the context of unimodular $F(R)$ gravity. As we demonstrated, the Newtonian potential is not affected in the case of unimodular $F(R)$ gravity, in contrast to the ordinary $F(R)$ gravity approach. Also we discussed which modes propagate in vacuum and the possibility a ghost mode appears. As we showed, the only propagating mode in vacuum is the graviton, and no ghost appears. Finally, the matter stability issue cannot be addressed by the usual technique used in ordinary $F(R)$ gravity, and we defer this study to a future publication. 

As we demonstrated in this paper, the unimodular $F(R)$ gravity formalism offers the possibility of realizing various bouncing cosmological scenarios which were exotic for the standard Einstein-Hilbert general relativity. Also an interesting extension of the unimodular gravity formalism to other modified gravity theories would be if we apply the formalism to theories with Lagrangian densities of the form $L= \sqrt{-g}\left(F(R, R^{\mu \nu}R_{\mu \nu} , R_{ \mu \nu \alpha \beta} R^{ \mu \nu \alpha \beta} \right)-\lambda)+\lambda$, or even simpler unimodular Gauss-Bonnet gravity with Lagrangian density of the form $L=\sqrt{-g}( F(G)-\lambda)+\lambda $. In addition, the same formalism could be applied to non-local gravity models with Lagrangian density of the form $L=\sqrt(-g)\left(F(R,R\square^m R,\square^d R)-\lambda\right)+\lambda $, with $m,n$ positive or negative integers. Furthermore, since every modified gravity theory is eventually tested by the potentiality of the theory to consistently describe relativistic objects, it is compelling to investigate if there are any relativistic star or black holes solutions of unimodular $F(R)$ gravity. Finally, the unimodular $F(R)$ gravity formalism should be investigated in the context of quantum vacuum corrections, or alternatively, to be embedded in the context of Loop Quantum Gravity \cite{Ashtekar:2011ni,Ashtekar:2007tv,Ashtekar:2011ni,Corichi:2009pp,Singh:2009mz,Bojowald:2008ik}, or any other potentially appealing quantum gravity theory. We hope to address some of these issues in a future work.

\section*{Acknowledgments}

This work is supported by MINECO (Spain), project 
FIS2013-44881 and I-LINK 1019 (S.D.O), by JSPS fellowship ID No.:S15127 (S.D.O.) and by Min. of Education and
Science of Russia
(S.D.O
and V.K.O) and  (in part) by
MEXT KAKENHI Grant-in-Aid for Scientific Research on Innovative Areas ``Cosmic
Acceleration''  (No. 15H05890) and the JSPS Grant-in-Aid for Scientific
Research (C) \# 23540296 (S.N.).

\section*{Appendix: Explicit Form of Some Parameters.}

In this Appendix we present the explicit form of various parameters appearing in Table \ref{newtable}. We start of with the parameters $\mathcal{A}_1$ and $\mathcal{A}_2$ the detailed form of which is,
\begin{align}
\label{mpalampala}
\mathcal{A}_1= & 2^{\frac{23}{18}-\frac{\sqrt{73}}{18}} 3^{-\frac{23}{18}+\frac{\sqrt{73}}{18}+\frac{1}{2} \left(-\frac{23}{18}+\frac{\sqrt{73}}{18}\right)} \left(\frac{69 C_2}{76}+\frac{3 \sqrt{73} C_2}{76}\right)(\rho_c)^{\frac{23}{18}-\frac{\sqrt{73}}{18}} \, , \nn
\mathcal{A}_2= & 2^{\frac{23}{18}+\frac{\sqrt{73}}{18}} 3^{-\frac{23}{18}-\frac{\sqrt{73}}{18}+\frac{1}{2} \left(-\frac{23}{18}-\frac{\sqrt{73}}{18}\right)} \left(-\frac{3 \sqrt{73}}{76}+\frac{69 C_1}{76}\right)(\rho_c)^{\frac{23}{18}+\frac{\sqrt{73}}{18}}\, ,
\end{align}
where $C_1$ and $C_2$ are arbitrary integration parameters. In addition, the parameters $n_1$ and $m_1$ are equal to,
\begin{equation}
\label{m1n1m2n2}
n_1=\frac{1}{2} i \left(i+\sqrt{71}\right)\, , \quad m_1=i \sqrt{15}\, .
\end{equation}
Also, the parameters $\mathcal{A}_3$ and $\mathcal{A}_4$ are equal to,
\begin{align}
\label{parametersa3a4}
\mathcal{A}_3 = & \frac{1}{2}\left( \frac{C_1\, 2^{\frac{1}{2}-\frac{i \sqrt{71}}{2}}\,\, _2F_1 \left(\frac{1}{2}-\frac{i \sqrt{71}}{2},\frac{1}{2}-i \sqrt{15}-\frac{i \sqrt{71}}{2},1-i \sqrt{15},-1 \right)}{\Gamma (1-i \sqrt{15})} \right. \nn
& \left. 
+\frac{2^{-1+i \sqrt{15}}\, C_2\,\sqrt{\pi}\,\,\, \Gamma \left(\frac{1}{2} \left(1+i \sqrt{15}+\frac{1}{2} i \left(i+\sqrt{71}\right)\right) \right)\sin \left(\frac{1}{2} \left(i \sqrt{15}+\frac{1}{2} i \left(i+\sqrt{71}\right)\right) \pi \right)}{ \Gamma (\frac{1}{2} \left(2-i \sqrt{15}+\frac{1}{2} i \left(i+\sqrt{71}\right)\right))} \right)\, , \nn
\mathcal{A}_4= &\frac{i 2^{-1+i \sqrt{15}}\, C_2\,\sqrt{\pi } \sqrt{\rho_c}\,\cos \left(\left(\frac{1}{2} \left(i \sqrt{15}+\frac{1}{2} i \left(i+\sqrt{71}\right)\right) \pi \right) \right)\,\Gamma \left(1+\frac{1}{2} \left(i \sqrt{15}+\frac{1}{2} i \left(i+\sqrt{71}\right)\right) \right)}{\Gamma \left(\frac{1}{2} \left(1-i \sqrt{15}+\frac{1}{2} i \left(i+\sqrt{71}\right)\right) \right)} \nn
& -\frac{2^{-\frac{1}{2}-\frac{i \sqrt{71}}{2}} \,\sqrt{15}\, C_1 \sqrt{\rho_c}\,\, _2F_1 \left(\frac{1}{2}-\frac{i \sqrt{71}}{2},\frac{1}{2}-i \sqrt{15}-\frac{i \sqrt{71}}{2},1-i \sqrt{15},-1 \right)}{\Gamma \left(1-i \sqrt{15}\right)} \nn
& -\frac{9 i 2^{\frac{1}{2}-\frac{i \sqrt{71}}{2}}\, C_1\,\sqrt{\rho_c}\,\, _2F_1(\frac{3}{2}-\frac{i \sqrt{71}}{2},\frac{1}{2}-i \sqrt{15}-\frac{i \sqrt{71}}{2},2-i \sqrt{15},-1)}{ \Gamma (2-i \sqrt{15})} \, ,
\end{align}
with $C_1$ and $C_2$ being arbitrary integration parameters. In addition, the parameters $\mathcal{A}_5$ and $\mathcal{A}_6$ are equal to,
\begin{align}
\label{parametersa3a4newpage}
\mathcal{A}_5=& C_2 \left(2^{1+\frac{6}{c^2}} 3^{\frac{1}{2}+\frac{3}{c^2}} \left(\frac{\left(1+\frac{6}{c^2}\right)^{\frac{12}{6+c^2}} \left(-4+c^2\right)}{\left(6+c^2\right)^2}\right)^{\frac{1}{2}+\frac{3}{c^2}}\right)^{\frac{2+c^2-\sqrt{4+20 c^2+c^4}}{12+2 c^2}}\, ,\nn
\mathcal{A}_6=&C_1 \left(2^{1+\frac{6}{c^2}} 3^{\frac{1}{2}+\frac{3}{c^2}} \left(\frac{\left(1+\frac{6}{c^2}\right)^{\frac{12}{6+c^2}} \left(-4+c^2\right)}{\left(6+c^2\right)^2}\right)^{\frac{1}{2}+\frac{3}{c^2}}\right)^{\frac{2+c^2+\sqrt{4+20 c^2+c^4}}{12+2 c^2}}\, ,
\end{align}
with $C_1$ and $C_2$ being again arbitrary integration parameters. Finally, the parameters $\Omega_1$ and $\Omega_2$ are,
\begin{equation}\label{superbounceomegas}
\Omega_1=\left(-\frac{\mathcal{A}_5}{-4+c^2}+\frac{3 \mathcal{A}_5 c^2}{2 \left(-4+c^2\right)}-\frac{ \mathcal{A}_2 \sqrt{4+20 c^2+c^4}}{2 \left(-4+c^2\right)}\right),\,\,\,\Omega_2=\left(-\frac{\mathcal{A}_6}{-4+c^2}+\frac{3 \mathcal{A}_6 c^2}{2 \left(-4+c^2\right)}+\frac{\mathcal{A}_1 \sqrt{4+20 c^2+c^4} }{2 \left(-4+c^2\right)}\right)\, .
\end{equation}

\end{document}